\documentstyle[preprint,aps]{revtex}
\begin{document}
\draft
\title
{Four-wave mixing and terahertz emission from three-level\\
systems in quantum wells: effects of inhomogeneous broadening}
\author{ Xuejun Zhu, Mark S. Hybertsen, P. B. Littlewood}
\address{
AT\&T Bell Laboratories, Murray Hill, New Jersey 07974
}
\author{Martin C. Nuss}
\address{
AT\&T Bell Laboratories, Holmdel, New Jersey 07733
}
\maketitle

\begin{abstract}

Using a perturbation theory based on
the density-matrix formulation,
we study the nonlinear optical responses of a noninteracting
three-level model system
to consecutive
coherent pulsed excitations, as realized in several recent
experiments involving
excitonic transitions in quantum wells.
The terahertz emission, which is a second order
response, and the four-wave mixing signal, a third order response, are
calculated within the rotating-wave approximation, in the presence of
detuning, dephasing, and inhomogeneous broadening.
We study the quantum beats in the photon echo
from transient four-wave mixing experiments of a three-level system.
We find that the temporal profile of the photon echo signal in the
four-wave mixing experiments depends very strongly on the
amount of inhomogeneous broadening of the energy
levels involved. Both the
position and the intensity of the peaks exhibit a
smooth evolution from strong quantum beats to a conventional
photon echo as the inhomogeneous broadening increases.
These features from our noninteracting model system are compared
with a recent experiment, and found to account for a
number of experimental observations.

\end{abstract}

\newpage
\section{Introduction}

The use of the ultra-short pulsed lasers ($e.g.,~Ti:Sapphire$)
has allowed
studies of a variety of coherent phenomena resulting
from the transient nonlinear interaction between the laser
field and a semiconductor or a semiconductor nanostructure at a time scale
of 100 $fs$ or even shorter
\cite{Shen84,Leo90hl,Leo91dw,Nuss92dw,Nuss92hl,Nuss93dw,Luo93dw,Koch92hl,Koch93mqw,Echoes,EchoesB,Kochetal,DSKim,NussRev}.
A simple two-level model system
\cite{YT79} has been used extensively in understanding
the observations from a variety of experiments.
When one of the two levels involved in the
optical transitions is composed
of two closely-spaced energy levels, various beating phenomena
are expected to occur at the frequency corresponding to the energy difference
between these two states nearby in energy.
One such example is a three-level system
comprising the heavy-hole, the light-hole, and the conduction
electron in an $AlGaAs/GaAs/AlGaAs$ quantum well
\cite{Leo90hl,Nuss92hl,Koch92hl}.
Another
example is composed of two quantum wells coupled to each other
through a
tunneling barrier \cite{Leo91dw,Nuss92dw,Nuss93dw,Luo93dw}.
In this case, some asymmetry between the two quantum wells is
often introduced, either by application of an electric field or
by design during sample growth, to increase the dipole moments
of the
desired optical transitions.
Several other possibilities have also been proposed and
explored in recent experiments \cite{Echoes}. Work has also been
extended to cases involving a strong magnetic field \cite{EchoesB}.

The experimental observations involving the
transient nonlinear optical responses of these three-level
systems include the
quantum beats (QB)
in the four-wave mixing (FWM) experiments, and the
coherent radiation at the difference frequency
between the two nearby levels
\cite{Leo90hl,Leo91dw,Nuss92dw,Nuss92hl,Nuss93dw,Koch92hl,Koch93mqw,Echoes}.
The FWM signal is a third-order response, and the difference frequency
radiation
is a second order response.
The energy spacing between the heavy-hole and the light-hole,
or between the coupled quantum well states is
a few $meV$ in these experiments. Therefore,
the beating phenomena of the FWM signal
occur with a period on the order of a picosecond; and
the radiation at the difference frequency occurs at
the terahertz range.

In this paper, we study these two nonlinear optical properties
with a noninteracting, three-level model system, taking into
account detuning, dephasing, and inhomogeneous broadening.
The density-matrix formulation and the
rotating-wave approximation \cite{Shen84,Kochetal,YT79}
are employed.
We use the
phenomenological
semiconductor optical Bloch equations to describe the interaction
of a semiconductor with the laser field \cite{Shen84,Kochetal}.
These equations can be derived microscopically,
and many-body interaction effects can be included
approximately \cite{Kochetal,DSKim}. In this paper,
we consider the simpler noninteracting model, ameanable to
analytic treatments,
seeking to study
phenomenologically the qualitative effects of detuning, dephasing,
and particularly inhomogeneous broadening.
A brief description of our work has been given previously \cite{ZHL}.
The inhomogeneous broadening effects have also been considered
by several other authors recently \cite{Cundiff}.

The balance of this paper is as follows. In Sec.~II, we
briefly review the two-level system solved
by the density-matrix formulation. We give results for
a three-level system for both the terahertz emission and
the four-wave mixing in Sec.~III, using the same perturbative
approach. Effects of finite dephasing and inhomogeneous
broadening are discussed in Sec.~IV for
the terahertz emission (Sec.~IVA) and for
the four-wave mixing (Sec.~IVB).
We conclude in Sec.~V.

\section{Density-Matrix Formulation and a Two-level System}

In order to gain some intuition into the problem and to establish
notation, we briefly
outline the solution to a two-level system first
considered by Yajima and Taira
\cite{YT79}, and in doing
so develop
a geometric interpretation for the response of the optical dipole
mement to the laser pulses \cite{Shen84}.

We label the two levels as $v$ for valence and $c$ for conduction
states. Only the relative energy matters and we denote it as
$\omega_0 = E_c - E_v$. (We take $\hbar = e = c = 1$ throughout.)
The ensemble-averaged density-matrix is defined as \cite{Shen84}:
\begin{equation}
\hat{n}_{\vec k} = \sum_{c,v} |\vec k><\vec k|
=
\left(\matrix{n_{v,\vec k} & \psi_{\vec k} \cr
              \psi^*_{\vec k} & n_{c,\vec k}}\right),
\label{dmdef}
\end{equation}
calculated at a time $t$. Here $\vec k$ may denote the internal
momentum of the electron-hole pair whose center-of-mass momentum
is zero in order to be optically active; or it may denote some internal
quantum number in the case of a bound exciton whose center-of-mass
momentum is once again zero.
We emphasize here that
we assume
there is no interaction between the excitons or the electron-hole
pairs of different $\vec k$'s. This assumption is presumably valid in the
limit of low excitation power.

The density-matrix follows a Liouville equation of motion
\cite{Shen84,Kochetal,YT79}:
\begin{equation}
{\partial \over {\partial t}}
\hat{n}_{\vec k} (t) = - i \biggl[ \hat{\varepsilon}_{\vec k} (t),
                                   \hat{n}_{\vec k} (t)
                           \biggr]
+
{\partial \over {\partial t}}
\hat{n}_{\vec k} (t) _{relax},
\label{EOM}
\end{equation}
where the Hamiltonian matrix is:
\begin{equation}
\hat{\varepsilon}_{\vec k} (t) = \left(\matrix{ 0 & -\mu_{\vec k} E(t) \cr
                                               -\mu_{\vec k}^*E^*(t) &
\omega_0}
                                 \right),
\label{Hmatrix}
\end{equation}
in which the off-diagonal terms represent the interaction with the
laser field, $\mu$ is the dipole moment between
states $|v>$ and $|c>$ projected in the direction of the applied
field $E(t)$ which is now taken as a scalar.
The relaxational term in the equation of motion is given by two
phenomenological dephasing times $T_1$ and $T_2$, often
referred to as the longitudinal
and transverse dephasing times respectively. Using these two
relaxation times, we may write:
\begin{equation}
{\partial \over {\partial t}}
\hat{n}_{\vec k} (t)_{relax}
= - \left(\matrix{ {1 \over T_1} (n_{v,\vec k} (t) - n_{v,\vec k}^0) &
                   {1 \over T_2} (\psi_{\vec k} (t) - \psi_{\vec k}^0) \cr
                 {1 \over T_2} (\psi^*_{\vec k} (t) -  \psi^{*0}_{\vec k} )&
                   {1 \over T_1} (n_{c,\vec k} (t) - n_{c,\vec k}^0)
                 } \right),
\label{EOMrelax}
\end{equation}
where the quantities with the superscript $^0$ denote the steady state
solution. In cases of transient excitation and transient response,
they are the same
as those before the laser pulses were introduced.
Physical quantities in this formalism are given by the trace of the
product of the operator matrix with the density-matrix. In particular
the dipole mement of the system is given by:
\begin{eqnarray}
p& =& Tr\biggl( \sum_{\vec k} \hat {p}_{\vec k} \hat {n}_{\vec k}
      \biggr) \nonumber \\
 & =& \sum_{\vec k} Tr \left(\matrix{0 & \mu_{\vec k} \cr
                                   \mu^*_{\vec k} & 0}\right)
                     \left(\matrix{n_{v,\vec k} & \psi_{\vec k} \cr
                                   \psi^*_{\vec k} & n_{c,\vec k}}\right)
  = 2\sum_{\vec k} Re(\mu_{\vec k} \psi^*_{\vec k}).
\label{dipolem}
\end{eqnarray}

For clarity, we will suppress the subscript $\vec k$ from
now on when
doing so causes no ambiguity. We also assume the $\vec k$-dependence
of the dipole matrix elements to be negligible, and therefore denote them
all by $\mu$. For the dipole moments between
the valence and the conduction bands,
this should be an excellent approximation
in considering low excitation power experiments with
electron-hole pair densities less than $10^{10} - 10^{11} cm^{-2}$ in
$GaAs$ quantum wells.
However, for the dipole moment between the heavy-hole and the light-hole
which we will introduce later for the three-level system, it is
a more questionable approximation that we have to make for
analytic solutions.

After we define:
\begin{eqnarray}
A_1 &=& \mu^*\psi, \\
A_2 &=& \mu\psi^*, \\
A_3 &=& n_c - n_v,
\end{eqnarray}
the application of the equation of motion yields:
\begin{eqnarray}
{{\partial A_1} \over {\partial t}} &=& i (\omega_0 A_1 - |\mu|^2 E A_3)
- {A_1-A_1^0 \over T_2}, \\
{{\partial A_2} \over {\partial t}} &=& i (- \omega_0 A_2 - |\mu|^2 E^* A_3)
- {A_2-A_2^0 \over T_2}, \\
{{\partial A_3} \over {\partial t}} &=& i (-2 E^* A_1 + 2 E A_2)
-{A_3-A_3^0 \over T_1}.
\end{eqnarray}

Within the rotating-wave approximation, we only consider the
resonant responses. We then write:
\begin{eqnarray}
E &=& e e^{i\omega t},\\
A_1 &=& a_1 e^{i\omega t},\\
A_2 &=& a_2 e^{-i\omega t},
\end{eqnarray}
with a detuning frequency
\begin{equation}
\omega_d = \omega - \omega_0.
\end{equation}

We now define a vector $\vec M = (M_1, M_2, M_3)$, in terms of
quantities $A_1~({\rm or}~a_1),~A_2~(a_2),$ and $ A_3$, whose
three components are given by:
\begin{eqnarray}
M_1 &=& a_1 + a_2,\\
M_2 &=& i( a_1 - a_2),\\
M_3 &=& |\mu| A_3.
\end{eqnarray}
It is then straightforward to find that the vector
$\vec M$ satisfies
the following equation of motion:
\begin{equation}
{{\partial \vec M} \over {\partial t}}
= \vec \Omega \times \vec M,
\end{equation}
with the vector $\vec \Omega$ given by:
\begin{equation}
\vec \Omega = ( - e_+ |\mu|, - e_- |\mu|, \omega_d),
\end{equation}
where we have used $e_+ = e + e^*$ and $e_- = i (e - e^*)$, and
for illustration we have taken $T_1$ and $T_2$ both equal to $\infty$.
Therefore, the thus-defined pseudo-vector $\vec M$ follows an
equation of motion that can be interpreted as a rotation around
the axis given by the unit vector $\vec \Omega / \Omega$ with an
angular velocity $\Omega$. This gives a geometric picture in the
rotating frame for the interaction of the dipole moment with the laser
field \cite{Shen84}.

Following Ref.~\cite{YT79}, the condition for photon echo
in FWM can be determined easily.
Let us consider cases where the time envelope-function of the
$E$-field $e$ is composed of two short pulses
$e_1$ and $e_2$, whose
durations are much shorter than the dephasing times $T_1$ and
$T_2$,
and also much shorter than
the inverse of the detuning frequency $\omega_d$, but
much longer than the inverse of $\omega_0$ or $\omega$. That is, for the
duration of the laser pulse, we can ignore the rotation of
the pseudo-vector $\vec M$ due to detuning, or the decrease of its
magnitude due to dephasing.
Under these conditions, the semiconductor photon echo can
be understood in the same way as the spin echo in
magnetic materials \cite{Shen84}. For the present two-level
system, the echo signal that is proportinal to
$e_2e_2e_1^*$ occurs
at a time:
\begin{equation}
t = 2 \tau,
\label{echot}
\end{equation}
where $\tau$ is the time-delay between the two consecutive
laser pulses, and the origin of the time is chosen at the arrival
of the first pulse.

\section{Solutions to a three-level system: terahertz emission
and four-wave mixing responses}

We now consider a three-level system using the same formalism
and the same notations as
given in the last section.
The energy levels are shown schematically in Fig.~1. The equation of motion
for the density-matrix bears the same form as Eq.~\ref{EOM},
with the Hamiltonian matrix now given by:
\begin{equation}
\hat{\varepsilon} = \left(
\matrix{ - \Delta E / 2 & -\mu_{12} E & - \mu_{13} E \cr
                 - \mu^*_{12} E^* & \Delta E/ 2 & - \mu_{23} E \cr
                 - \mu^*_{13}E^* & -\mu_{23}^* E^* & \omega_0
       }
                    \right),
\label{3levelH}
\end{equation}
where $\mu_{ij}$'s are the dipole matrix elements between levels $i$ and
$j$, $\Delta E$ is the energy spacing between level 1 and level 2, and
$\omega_0$ is the energy spacing between the center of levels 1 and 2
and level 3, and $\omega_0 \gg \Delta E$.
The response of the system
naturally separates into two terms according to the
dominant time scales:
\begin{equation}
p_{THz} = 2Re(\mu_{12}\psi_{12}^*),
\label{pthzdef}
\end{equation}
and
\begin{equation}
p_{FWM} = 2Re(\mu_{13}\psi_{13}^*) +
          2Re(\mu_{23}\psi_{23}^*),
\label{pfwmdef}
\end{equation}
where $p_{THz}$ is the dipole moment that gives
rise to terahertz radiation for $\Delta E \sim meV$,
and $p_{FWM}$ is the dipole moment which contains
information on the FWM signal at the excitation frequency
$\omega$.

Using the rotating-wave approximation, we define:
\begin{eqnarray}
E &=& e(t) e^{i\omega t},\\
\psi_{13} &=& f_{13}(t) e^{i\omega t},\\
\psi_{23} &=& f_{23}(t) e^{i\omega t},\\
\psi_{12} &=& f_{12}(t).
\end{eqnarray}
{}From the equation of motion for the three-level
system, we find that the off-diagonal elements of the
density-matrix satisfy:
\begin{eqnarray}
{{\partial f_{12}} \over {\partial t}} + {f_{12} \over T_{12}}
&=& i\Delta E f_{12} - i \mu^*_{23} e^* f_{13}
                   + i \mu_{13}   e   f_{23}^*,\\
{{\partial f_{13}} \over {\partial t}} + {f_{13} \over T_{13}}
&=& i(\omega_0 + \Delta E/2 - \omega)  f_{13} - i \mu_{13} e (n_{11} - n_{33})
                   - i \mu_{23} e f_{13},\\
{{\partial f_{23}} \over {\partial t}} + {f_{23} \over T_{23}}
&=& i(\omega_0 - \Delta E/2 - \omega)  f_{23} - i \mu_{23} e (n_{21} - n_{33})
                                            - i \mu_{13} e f_{12}^*,
\end{eqnarray}
and the diagonal matrix elements, or the populations, satisfy:
\begin{eqnarray}
{{\partial n_{11}} \over {\partial t}} + {n_{11} \over T_{11}}
&=&   i \mu_{13} e f_{13}^*
  - i \mu_{13}^* e^* f_{13},\\
{{\partial n_{22}} \over {\partial t}} + {n_{22} \over T_{22}}
&=&   i \mu_{23} e f_{23}^*
  - i \mu_{23}^* e^* f_{23},\\
{{\partial n_{33}} \over {\partial t}} + {n_{33} \over T_{33}}
&=&
-\bigl({{\partial n_{11}} \over {\partial t}} + {n_{11} \over T_{11}}
 \bigr)
-\bigl({{\partial n_{22}} \over {\partial t}} + {n_{22} \over T_{22}}
 \bigr).
\end{eqnarray}
We have assumed $|\mu_{12}| \ll |\mu_{13}|, |\mu_{23}|$.

Note that these are all linear differential equations, and thus
the solutions to them are additive in terms of the source terms on
their right-hand-sides. This is a great advantage and makes
it a much simpler problem to deal with.
The linearity of these equations
does not hold in general
once interaction effects are included \cite{Kochetal}.

We now consider the situation where the second fastest
time scale in the problem is the laser pulse duration $t_L$, {\it i.e.\/},
$t_L \ll 1/\Delta E, T_{ij}$; it is also much shorter than the
time-delay between the laser pulses $\tau$.
In this case, we can approximate the envelope-function
$e(t)$ by $\delta$-functions. In particular, in the case of
two phase-locked short laser pulses, we may write:
\begin{equation}
e(t) = e_1 \delta (t) + e_2 \delta (t-\tau),
\label{twopulses}
\end{equation}
where the phases of $e_1$ and $e_2$ differ by a specifiable amount
that is experimentally tunable \cite{Nuss93dw,Luo93dw}.

Equations 29-34 are solved perturbatively in the strength
of the laser field $e$ in Eq. 35. Thus our solutions are only applicable
below the saturation intensity \cite{Shen84}.
The initial conditions and
thus the zeroth-order solutions are:
\begin{eqnarray}
(n_{11} - n_{33} ) &=& 1, \\
(n_{22} - n_{33} ) &=& 1, \\
f_{ij}             &=& 0.
\end{eqnarray}

To second order in $e$, we have for the polulations:
\begin{eqnarray}
n_{11}^{(2)} - 1 &=&- |\mu _{13}|^2 \biggl(
 e^{- t/T_{11}} \theta (t) |e_1|^2 \nonumber \\
 &&+e^{- (t-\tau)/T_{11}} \theta (t-\tau) |e_2|^2 \nonumber \\
 &&+e^{- (t-\tau)/T_{11}} \theta (t-\tau) \bigl(
 e_1^*e_2 e^{-i(\omega_0 + \Delta E/2 -\omega)\tau - \tau/T_{13}}
+e_1e_2^* e^{ i(\omega_0 + \Delta E/2 -\omega)\tau - \tau/T_{13}}
                                       \bigr)
                                   \biggr)
,
\label{n11}
\end{eqnarray}
\begin{eqnarray}
n_{22}^{(2)} - 1 &=& - |\mu _{23}|^2 \biggl(
 e^{- t/T_{22}} \theta (t) |e_1|^2  \nonumber \\
&&+e^{- (t-\tau)/T_{22}} \theta (t-\tau) |e_2|^2  \nonumber \\
&&+e^{- (t-\tau)/T_{22}} \theta (t-\tau) \bigl(
 e_1^*e_2 e^{-i(\omega_0 - \Delta E/2 -\omega)\tau - \tau/T_{23}}
+e_1e_2^* e^{ i(\omega_0 - \Delta E/2 -\omega)\tau - \tau/T_{23}}
                                       \bigr)
                                   \biggr)
,
\label{n22}
\end{eqnarray}
\begin{eqnarray}
n_{33}^{(2)} - 0 &= & |\mu _{13}|^2 \biggl(
 e^{- t/T_{33}} \theta (t) |e_1|^2  \nonumber \\
&&+e^{- (t-\tau)/T_{33}} \theta (t-\tau) |e_2|^2  \nonumber \\
&&+e^{- (t-\tau)/T_{33}} \theta (t-\tau) \bigl(
 e_1^*e_2 e^{-i(\omega_0 + \Delta E/2 -\omega)\tau - \tau/T_{13}}
+e_1e_2^* e^{ i(\omega_0 + \Delta E/2 -\omega)\tau - \tau/T_{13}}
                                       \bigr)
                                   \biggr) \nonumber \\
&&                   +|\mu _{23}|^2 \biggl(
e^{- t/T_{33}} \theta (t) |e_1|^2  \nonumber \\
&&+e^{- (t-\tau)/T_{33}} \theta (t-\tau) |e_2|^2 \nonumber \\
&&+e^{- (t-\tau)/T_{33}} \theta (t-\tau) \bigl(
 e_1^*e_2 e^{-i(\omega_0 - \Delta E/2 -\omega)\tau - \tau/T_{23}}
+e_1e_2^* e^{ i(\omega_0 - \Delta E/2 -\omega)\tau - \tau/T_{23}}
                                       \bigr)
                                   \biggr)
,
\label{n33}
\end{eqnarray}
and for the dipole moment responsible for the terahertz emission:
\begin{eqnarray}
\mu_{12}^*f^{(2)}_{12}
&=& - \mu_{12}^*\mu _{23}^*\mu_{13}
                                   \biggl(
 |e_1|^2  \theta (t) e^{(i\Delta E-1/T_{12})t} \nonumber \\
&&+|e_2|^2  \theta (t-\tau) e^{(i\Delta E-1/T_{12})(t-\tau)}\nonumber \\
&&+         \theta (t-\tau) e^{(i\Delta E-1/T_{12})(t-\tau)}
\bigl(
 e_1e_2^* e^{ i(\omega_0 + \Delta E/2 -\omega)\tau - \tau/T_{13}}
+e_1^*e_2 e^{-i(\omega_0 - \Delta E/2 -\omega)\tau - \tau/T_{23}}
\bigr)
                                   \biggr)